# OBTAIN: Real-Time Beat Tracking in Audio Signals

Ali Mottaghi, Kayhan Behdin, Ashkan Esmaeili, Mohammadreza Heydari, and Farokh Marvasti

Sharif University of Technology, Electrical Engineering Department, and

Advanced Communications Research Institute (ACRI), Tehran, Iran

Email: aesmaeili@stanford.edu

*Abstract*—In this paper, we design a system in order to perform the real-time beat tracking for an audio signal. We use Onset Strength Signal (OSS) to detect the onsets and estimate the tempos. Then, we form Cumulative Beat Strength Signal (CBSS) by taking advantage of OSS and estimated tempos. Next, we perform peak detection by extracting the periodic sequence of beats among all CBSS peaks. In simulations, we can see that our proposed algorithm, Online Beat TrAckINg (OBTAIN), outperforms state-of-art results in terms of prediction accuracy while maintaining comparable and practical computational complexity. The real-time performance is tractable visually as illustrated in the simulations. [1]

## I. INTRODUCTION

The beat is a salient periodicity in a music signal. It provides a fundamental unit of time and foundation for the temporal structure of the music. The significance of beat tracking is that it underlies music information retrieval research and provides for beat synchronous analysis of music. It has applications in segmentation of audio, interactive musical accompaniment, cover-song detection, music similarity, chord estimation, and music transcription. It is a fundamental signal processing task of interest to any company providing information services related to music [1].

### A. Related Works

Many works have been carried out in offline beat tracking. One can find effective algorithms in the literature which perform beat tracking in an offline fashion [2]. It is however important to mention some of previous works on beat tracking. Aubio [3] is a real-time beat tracking algorithm which is available as a free application. Aubio has been used in entertainment applications like Sonic Runway [4] in 2016. IBT [5] is a state-of-the-art real-time algorithm in this field. IBT is based on BeatRoot [6] tracking strategy which is a state-of-the-art offline tracking algorithm. BeatRoot system takes advantage of two pre-processing and processing stages. In the pre-processing stage, a time-domain onset detection algorithm is used which calculates onset times from peaks in the slope of the amplitude envelope. The processing stage consists of two blocks. The first block uses a clustering algorithm on inter-onset intervals and generates a set of tempo hypotheses by examining the relationships between clusters. The second block is a tracking block. In this block, a multiple agent architecture is used, where each agent represents a hypothesis about the tempo. Performance of each agent due to the data is evaluated and the agent with the best performance returns output of the tracking system. Another state-of-the-art algorithm is Ellis [2]. Although Ellis method is not causal, some blocks of our system are based on this method.

### B. Our Contributions

The goal of this paper is to provide a fast and competitive beat tracking algorithm for audio signals that can be easily implemented in real-time setting. As our key contributions, we

1) propose a simple yet fast beat tracking algorithm for audio signals,
2) extend the algorithm to real-time implementation,
3) compare the algorithm to previous results to show that it outperforms state-of-art algorithm prediction accuracy while maintaining comparable and practical computational complexity,
4) implemented our method on an embedded system (Raspberry Pi 3) to demonstrate its effectiveness and reliability in real-time beat tracking,
5) participated in a real-world challenge (IEEE SP Cup 2017) and received honorable mention for our excellent beat tracking algorithm and annotation.

## II. OBTAIN ALGORITHM

The proposed approach follows a relatively common architecture with the explicit design and tries to simplify each step and modify them. Therefore, they can be applied in the real-time setting. We call our algorithm OBTAIN (a pseudo-abbreviation of Online Beat TrAckINg). We will elaborate upon the blocks of this system throughout the paper and compare it to state-of- art methods. There are four main stages to the algorithm, as shown in Fig. 1. The initial audio input has a sampling rate of 44100 Hz.

### A. Generating Onset Strength Signal (OSS)

Beat tracking is an audio signal processing tool which is based on onset detection. Onset detection is an important issue in signal processing. It can be widely seen in different pieces of research that onset detection is used such as music signal processing [7], neural signal processing (EEG, ECoG, and FMRI), and other biomedical signal processing areas such as electro-cardiac signals to name but a few [8], [9]. In musical signal processing, there would be many practical cases where this onset detection would prove to be important. Visual effects in musical applications may work based on real-time onset detection as in music player applications. The purpose is to capture abrupt changes in the signal at the beginning of the transient region of notes [7].

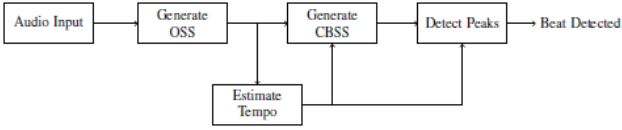

Figure 1. Block diagram of OBTAIN

Since onset detection is a basic part of many audio signal analysis, many algorithms are implemented for this purpose and most of them can be applied to real-time setting like introduced method in [7]. This is one of latest methods proposed for this issue. We split the subject audio file into overlapping windows. In order to detect onsets, we require to perform our algorithm on a sequence of samples since working with one sample at a time we cannot derive any onset. We require processing on a frame of samples to implement Fast Fourier Transform (FFT) in order to have access to an array of samples to learn the pattern of beats. Therefore, we consider windows of samples where each window is of size 1024 samples, i.e. we suppose each window contains 1024 samples. Thus, the sampling rate is $\frac{44100}{1024} = 43.06 Hz$. We also consider the overlapping ratio equal to $87.5\%$. In other words, we choose the Hop size ($H$) parameter equal to 128 samples and consider the overlapping ratio of the new input series of samples with the stacked frame equal to 87.5%. The reason behind choosing large overlap is to enhance accuracy. If we try to maintain the structure of a specific frame for several stages, the efficiency of the algorithm performance increases since the desired frame stays somehow in the memory for a while. This is, in fact, equivalent to 87.5% decrease in sampling rate. Then, we compute FFT of each window. We normalize the data by dividing the components to a normalizing value. This normalizing value is chosen as follows: We consider a fixed span of the frequency band at the beginning of FFT of the audio signal and find the maximum absolute value in this span of time. We suppose this is a good approximate of the maximum component for the entire frequency range. Afterward, threshold the components below an empirical level $74dB$ to zero (An empirical noise level cancellation) [10]. Next, we apply log-compression on the resulted window. The log compression is carried out as follows: Let $X$ denote the resulted window, then the log compressed signal is:

$$\Gamma_\lambda(X) = \frac{\log(1+\gamma|X|)}{\log(1+\gamma)} \quad (1)$$

It is worth noting that after log-compressing we perform a further normalization step in order to be assured the maximum of the signal is set to 1. Log compression is carried out in order to reduce the dynamic range of the signal and adapt the resulted signal to the human hearing mechanism which is logarithmically sensitive to the voice amplitude [7].

We define $Flux$ function based on the log-compressed signal spectrum $\Gamma_\lambda$ as follows:

$$Flux[n] = \sum_{k=0}^{K} |\Gamma_\lambda[n+1,k] - \Gamma_\lambda[n,k]|_+ \quad (2)$$

where $|x|_+$ is $\max\{x, 0\}$.

This function is, in fact, discrete temporal derivative of the compressed spectrum [7]. Now we apply a Hamming window ($h[n]$) of length $L = 15$ with the cutoff frequency equal to $7Hz$ in order to remove noise components from the OSS. OSS can be derived by applying the Hamming filter on the $Flux$ as follows:

$$OSS[n] = \sum_{k=n-\left[\frac{L}{2}\right]}^{n+\left[\frac{L}{2}\right]} Flux[k] \times h[k] \quad (3)$$

Fig. 2 shows OSS for an audio signal.

*B. Tempo Estimation*

We store the OSS we obtained in the previous phase into a buffer of length 256. Each buffer contains 256 OSS samples. Two intuitive reasons behind this choice of length for the buffer could be first, the robustness of real-time process, and second, the time required for the buffer to load enough samples for detection would be approximately 3 secs, which is compatible to human hearing capability in beat detection. [11]

We use the algorithm which was described in [11] as our baseline for tempo estimation, although it is an offline algorithm. This algorithm is based on cross-correlation with pulses. To estimate tempo, autocorrelation is applied to frames of the OSS. After enhancing harmonics of the autocorrelation, the 10 top peaks of the enhanced autocorrelation are chosen. These peaks should satisfy the maximum and minimum tempo limits which are related to time lags in the autocorrelation. The peaks of the autocorrelation are the candidate tempos. Once the candidate tempos are chosen, cross-correlation with ideal pulse trains is used to assign scores to the candidate tempos. Scoring is based on the highest and variance of cross-correlation values among all possible time shifts for pulse trains. The instance tempo evaluated from each frame of the OSS is the candidate tempo with the highest score. In the next step, we accumulate all instance tempos evaluated from frames of the OSS. More details on this method are available in [11]. Since our algorithm is causal, we only use instance tempo produced by frames of the OSS which have appeared until now, in contrary to the method proposed in [11] which uses the entire frames.

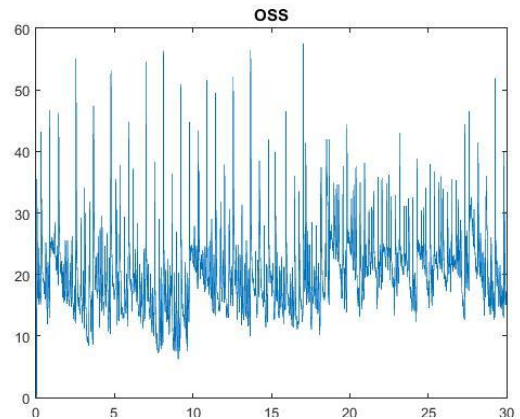

Figure 2. The OSS signal for audio signal no. 10 in dataset Open in [15].

An important improvement we obtained from our algorithm is that tempo variation becomes verifiable; In the sense that variations of audio are distinguishable from undesirable fluctuations. To comply with this subject of action, we have added another stage to this block. In the final stage, we store the history of tempo for about 7 last seconds of the music. In the first 7 seconds of the audio signal we just use overall estimation by accumulating tempos as described in [11]. Next, we compare the resulted accumulated tempo with the mean of the tempo history for each frame. If tempos differ significantly (more than 5 BPM), we use the mean value because tempo fluctuations result in asynchronies in the blocks using tempo if resulted accumulated tempo is a harmonic of the mean of the tempo history. Thus, it is better not to change the tempo frequently. If this change is long-lasting and the new tempo is not a harmonic of the mean tempo, we finally change our tempo after about 1 second. It is noticeable that since each time the instance tempo is resulted from about 7 last seconds of music our choices sound reasonable.

### C. Cumulative Beat Strength Signal (CBSS)

At this stage, we want to score the frames according to their possibility of being selected as a beat. This can be done using CBSS which was first proposed by [2]. Here frames are actually our audio samples which they are selected by overlapping windows and should be determined if they a beat or not. To generate CBSS for each frame, we initially look for the previous beat which is observed as a peak in this signal. CBSS for a frame is equal to the weighted sum of OSS in the corresponding frame and the value of CBSS of the last beat using different weights. Now, we explain how to find the last beat. To this end, we employ a log-Gaussian window [12]. To specify the location of the beats, we use a recursive method to assign a score to each sample which determines the beat power in the working frame. The maximum value among these scores specify the beat location. CBSS is obtained via calculating the summation of two terms: one from the previous frames and the other is related to the current frame. Let $\tau_b$ be the estimated beat period from the tempo estimation. We consider a scoring window on the span $[n - \frac{\tau_b}{2}, n + \frac{\tau_b}{2}]$ for the $n$-th sapmle. Then, we form the log-Gaussian window as follows:

$$W[v] = e^{\left[-\left(\eta \log\left(-\frac{v}{\tau_b}\right)\right)^2/2\right]} \quad (4)$$

where $v \in [-\tau_b/2, -2\tau_b]$, and $\eta$ determines the log-Gaussian width. $CBSS[n]$ denotes the CBSS at each sample. $\Phi[n]$ is defined as follows:

$$\Phi[n] = \max_v W[v] CBSS[n + v] \quad (5)$$

This value approximately determines what the score of the previous beat was. Implementations agree with this assumption in assigning scores to the beats [2]. Finally, the score for each frame is calculated as follows:

$$CBSS[n] = (1 - \alpha)OSS[n] + \alpha\Phi[n] \quad (6)$$

This structure results in quasi-periodic CBSS. Therefore, even when the signal is idle, previous scores could be used to obtain the next beat. The periodic structure is improved throughout the learning process of the algorithm, and the estimation accuracy increases. The choice of $\alpha$ is carried out using cross-validation in several implementations on the training data set selected randomly from the main dataset (80% of the dataset).

### D. Beat Detection

At final stage of our algorithm, periodic peaks of CBSS are detected in a real-time fashion, and the output signal "beatDetected" is a flag which is set to 1 if a peak is detected in that frame. The method described in [2] uses a non-causal beat tracker which is not practical in real-time systems. Therefore, we use a more sophisticated system to overcome this issue.

This block takes advantage of two separate parallel systems to enhance reliability of the system performance. The initial system simply tracks periodic beats without considering the beat period, while the second system is totally dependent on the beat period. The main assumption is that if the beat period is not detected correctly, the cumulative signal still maintains its periodic pattern. Therefore, the second system is a correction system. The outcome of this block is based on the comparison of the CBSS values in the peak locations detected by these two systems. The system yielding higher average is chosen. Each frame consisting of 512 samples of the CBSS is fed into this block in a buffer. Two consecutive buffers are overlapped with 511 common samples (like FIFO concept). To reduce the complexity of computations, both systems do not function for all frames. The first system only functions when the distance between the current sample and the time the previous beat is detected falls in the span (BP-10, BP+7), i.e. the span we expect to observe a peak. This span is chosen since the beats must be detected within at most 0.1s, and further delay is not practical for a real-time system.

If no beat is detected in the mentioned span, the system finally turns the flag to 1 to maintain periodicity of the peak locations. The second system works exactly in the middle of the two beats, i.e. when the half of the BP is passed since the last detected beat and stops detecting till the next beat is detected. Thus, it must be stored in a buffer until the next beat is detected for comparison. The correction made by the second system is through this buffer. When the second system achieves a higher average of the CBSS values in the peak locations in comparison to the first system, the first system is corrected by considering the peaks detected by the second system as the previous beats, i.e. the last peak detected by the second system is considered as the last detected beat. Therefore, the next detected peaks will be continuation of the peak sequence with higher CBSS values which are more likely to be correct beats.

Now, we specify the mechanism of each system as follows:

- The main (initial system):

Here, we take advantage of the method introduced in [13]. The main concept is that the periodic beats have the largest value in comparison to the rest of samples within windows of length $\tau_b$; therefore, we initially look for the main BP and afterward look for the maximum values in windows of length BP. A summary of the method provided in [13] is summarized as follows: We assume that the input series to this system is the CBSS (whose peaks should be detected). We initially subtract the linear predictor of the data from the samples and denote the resulted signal with $x$. Afterwards, the $LMS \in \mathbb{R}^{\left\lceil \frac{N}{2}-1 \right\rceil \times N}$ is defined as follows:

$$m_{k,i} = \begin{cases} 0, & x_{i-1} > x_{i-k-1} \land x_{i-1} > x_{i+k-1} \\ 1+r, & otherwise \end{cases} \quad (7)$$

where $r$ is a uniform random number in $[0,1]$.

Then, the rows of the matrix $LMS$ are added together. Let $\gamma_k$ denote the result of summation for each column. Now, let $\lambda = argmin(\gamma_k)$. Clipping the matrix from the $\lambda$-th row, we will have the resulted $ScaledLMS$, finally the columns which have zero variance in the $ScaledLMS$ matrix determine the peak locations.

These evaluations are carried out for each frame. If the last sample of the frame is a peak, the flag "beatDetected" turns to 1.

- The second system:

In this system, a pulse train with the same period as BP is generated, and afterward, is cross-correlated with the CBSS.

$$CCor\,[n] = CBSS\,[n] * PulseTrain\,[-n] \quad (8)$$

The peak location of the resulted signal determines the displacement required for the pulse train to be matched with the CBSS peaks. Thus, the second system detects periodic peaks of the CBSS. If the peaks detected by this system have higher average CBSS values in comparison to the first system, the first system is tracking wrong peaks. Therefore, the first system is forced to track this new peak sequence.

III. SIMULATION RESULTS

A. Datasets

We have used two different datasets to show performance of our system. The first dataset is *Ballroom* dataset which is available in [14]. This dataset consists of 698 excerpts. Duration of each excerpt is about 30 seconds. The second dataset is ICASSP SP cup training dataset provided in [15]. This dataset consists of 50 selected excerpts. Duration of each excerpt is 30 seconds.

B. Evaluation Measures

We evaluate the performance of our method based on the four continuity based metrics defined in [16]. The four metrics are $CML_c, CML_t, AML_c, AML_t$ respectively. These four metrics are based on the continuity of a sequence of detected beats. $CML_c$ is the ratio of the longest continuously correctly tracked section to the length of the file, with beats at the correct metrical level. $CML_t$ is the total number of correct beats at the correct metrical level. $AML_t$ is the ratio of the longest continuously correctly tracked section to the length of the file, with beats at allowed metrical levels. $AML_c$ the total number of correct beats at allowed metrical levels.

We also evaluate our method based on the P-score metric and f-Measure as introduced in [17]. Let $b$ denote number of correctly detected beats and $p$ denote number of false detected beats and $n$ denote number of undetected beats. P-score and f-Measure are defined in (9) and (10).

$$P = \frac{b}{b + \max(p,n)} \quad (9)$$

$$F = \frac{b}{b + \frac{p+n}{2}} \quad (10)$$

We have fixed the Tempo tolerance to 17.5% and Phase tolerance to 25% for the four continuity based metrics. The tolerance window is set to 17.5% for f-Measure.

C. Results

Fig. 3 shows our algorithm's performance on beat detection for Audio number 10 in the [15]. Fig. 4 shows our method's performance on a difficult excerpt (Audio number 76 in SMC dataset [18]) with variable tempo. As it can be seen in Fig. 3, there is a "transient" state which lasts about 5 seconds. Because the tempo estimation block needs few seconds to estimate correct and stable tempo, this transient state is inevitable. Also, at the 5th second, the beat detection block decides to correct the first system by the procedure explained before. After this moment, the correct peak locations in the CBSS are chosen, in contrary to the first 5 seconds. Generally, this transition between systems can occur at any moment. It is worth noting that the CBSS peaks become shaper as the music continues.

To demonstrate our system's performance, we have compared our system to four other methods. These methods are Ellis method [2], Aubio [3], BeatRoot [6] and IBT [5]. Since our algorithm is partly based on Ellis, we have chosen this method for comparison. Also, IBT and BeatRoot are chosen as two state-of-the-art causal and non-causal methods. The results of comparison of methods are provided in tables I and II. Since BeatRoot and Ellis are non-causal, they are labeled as NC in tables.

The four metrics $CML_c, CML_t, AML_c, AML_t$ evaluated on dataset [15] for these methods are plotted versus the phase tolerance as provided in Fig. 5, 6, 7 and 8.

To verify the reliability of our method, we simulate these methods on two different scenarios. First, we modify excerpts from dataset [15] by adding white Gaussian noise to reduce $SNR$ level of excerpts to $15dB$. Results provided in table III are achieved by averaging results on 4 sets of independently modified data. It can be observed that our method is still our performing other causal methods. To simulate the effect of low sampling rate, we filter excerpts using a low pass filter with scaled cut off frequency of $\frac{4000}{44100}$. Therefore, this filter has a cut-off frequency of $2KHz$ in continuous domain. Results of simulation on these excerpts are also provided in table IV. We conclude from the provided results that our method outperforms other real-time methods. The main advantage of our system over Ellis method is our peak detection system. Causality and accuracy are two improvements we have obtained by our peak detection system. Also, in comparison to BeatRoot, our system maintains comparable performance and complexity.

## IV. Implementation

We also have implemented our method on an embedded system (Raspberry Pi 3) to check its effectiveness and reliability in real-time beat tracking. The algorithm developed in MATLAB Simulink is converted to a C/C++ code with the cooperation of the algorithm developer and software developer, to make an implementation of the algorithm from scratch. Simulink (and basically MATLAB) has the possibility to generate the software or hardware design that corresponds to block diagrams which can also contain M-file functions or scripts. We use this feature to generate the C code. After proper configuration of Simulink Code Generator and its solver, we are able to generate the C/C++ code that implements the exact same functionality. Basic mathematical operations and also some complex operations such as the FFT (which is the most complicated procedure in the algorithm) are directly performed in a plain source code, without using external libraries. Operations such as loading audio files and playback are implemented by connecting the generated application to MATLAB exclusive libraries. Our source codes, application and video of real time beat tracking can be accessed at [19] and high-quality version of our recorded video at [20].

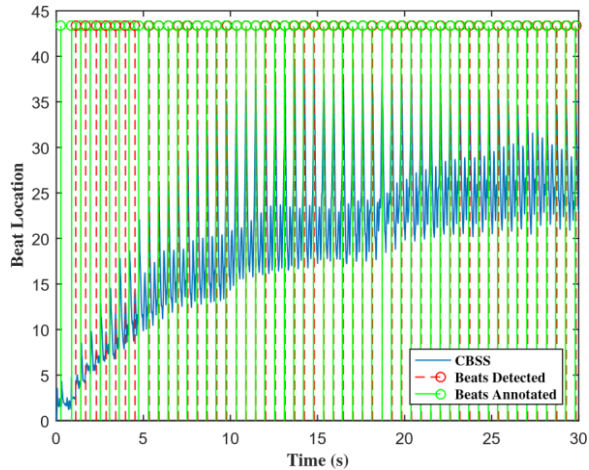

Figure 3. The $CBSS$ signal for a.5udio signal no. 10 in dataset Open in [15] and real-time beat detection.

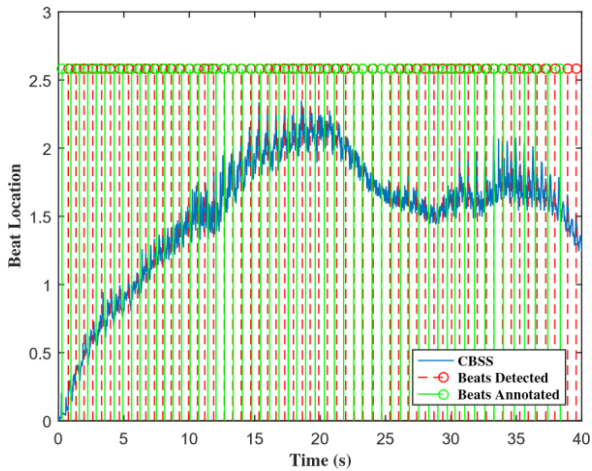

Figure 4. The $CBSS$ signal for difficult excerpt in dataset SMC in [18] and real-time beat detection.

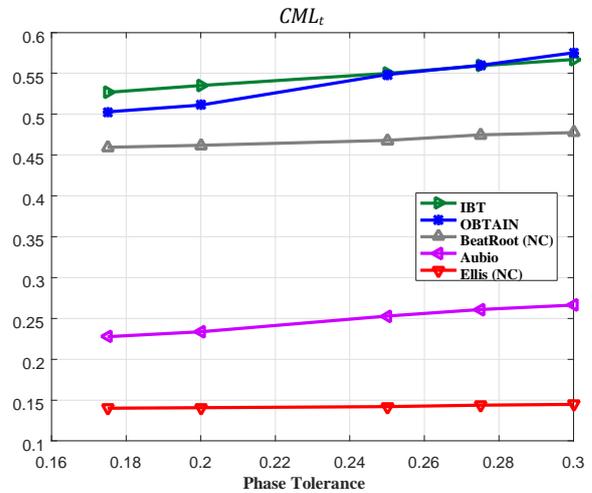

Figure 5. $CML_t$ vs. Phase tolerance.

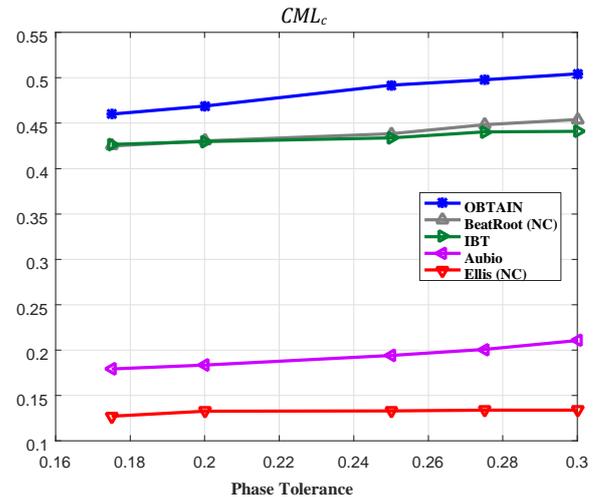

Figure 6. $CML_c$ vs. Phase tolerance

Table I. COMPARISON OF PERFORMANCES OF THE METHODS ON *Ballroom* (IN %).

| method | $AML_t$ | $AML_c$ | $CML_t$ | $CML_c$ | $P-score$ | $f-Measure$ | average |
|---|---|---|---|---|---|---|---|
| OBTAIN | 77.36 | 70.37 | 49.91 | 45.57 | 66.73 | 72.53 | 63.75 |
| IBT | 76.41 | 66.92 | 54.70 | 49.02 | 63.47 | 71.52 | 63.67 |
| Ellis (NC) | 83.20 | 77.02 | 31.53 | 29.21 | 65.81 | 74.21 | 60.16 |
| Aubio | 34.54 | 27.37 | 25.07 | 20.19 | 71.00 | 66.29 | 40.74 |
| BeatRoot (NC) | 87.41 | 77.77 | 59.10 | 54.65 | 77.81 | 83.97 | 73.45 |

Table II. COMPARISON OF PERFORMANCES OF THE METHODS ON IEEE DATASET (IN %).

| method | $AML_t$ | $AML_c$ | $CML_t$ | $CML_c$ | $P-score$ | $f-Measure$ | average |
|---|---|---|---|---|---|---|---|
| OBTAIN | 74.97 | 66.99 | 54.86 | 49.17 | 70.79 | 68.59 | 64.23 |
| IBT | 67.72 | 54.04 | 55.00 | 43.38 | 65.70 | 67.28 | 58.85 |
| Ellis (NC) | 72.99 | 64.77 | 14.20 | 13.31 | 62.58 | 65.58 | 48.91 |
| Aubio | 33.87 | 25.41 | 25.28 | 19.41 | 74.16 | 64.48 | 40.44 |
| BeatRoot (NC) | 82.03 | 72.88 | 46.79 | 43.85 | 72.87 | 74.74 | 65.53 |

Table III. COMPARISON OF PERFORMANCES OF THE METHODS ON IEEE DATASET (NOISY) (IN %).

| method | $AML_t$ | $AML_c$ | $CML_t$ | $CML_c$ | $P-score$ | $f-Measure$ | average |
|---|---|---|---|---|---|---|---|
| OBTAIN | 68.69 | 59.30 | 50.00 | 43.87 | 67.79 | 64.36 | 59.00 |
| IBT | 67.06 | 53.59 | 54.19 | 43.16 | 66.11 | 67.20 | 58.55 |
| Ellis (NC) | 69.22 | 60.69 | 9.01 | 8.76 | 62.61 | 63.93 | 45.70 |
| Aubio | 33.77 | 23.04 | 23.89 | 16.62 | 73.85 | 64.85 | 39.34 |
| BeatRoot (NC) | 80.35 | 68.81 | 43.40 | 38.31 | 72.73 | 74.64 | 63.04 |

Table IV. COMPARISON OF PERFORMANCES OF THE METHODS ON IEEE DATASET (FILTERED) (IN %).

| method | $AML_t$ | $AML_c$ | $CML_t$ | $CML_c$ | $P-score$ | $f-Measure$ | average |
|---|---|---|---|---|---|---|---|
| OBTAIN | 69.51 | 56.01 | 58.61 | 48.13 | 75.55 | 70.90 | 63.12 |
| IBT | 60.66 | 44.90 | 48.72 | 35.44 | 65.39 | 65.42 | 53.42 |
| Ellis (NC) | 74.72 | 64.66 | 9.43 | 8.82 | 60.82 | 64.03 | 47.08 |
| Aubio | 29.37 | 21.06 | 21.66 | 16.65 | 73.28 | 61.93 | 37.33 |
| BeatRoot (NC) | 73.98 | 61.82 | 39.67 | 35.11 | 72.68 | 72.83 | 59.35 |

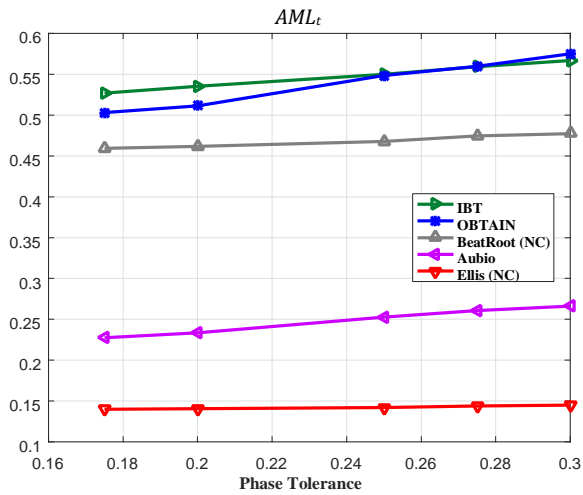

Figure 7. $AML_t$ vs. Phase tolerance

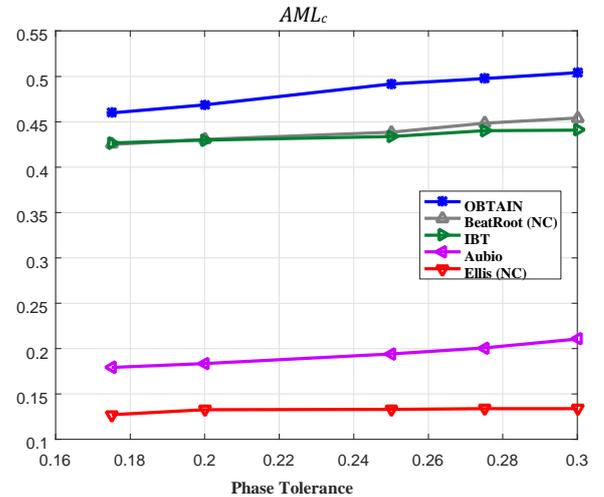

Figure 8. $AML_c$ vs. Phase tolerance.

## V. CONCLUSION

In this paper, we propose an algorithm towards real-time beat tracking (OBTAIN). We use OSS to detect onsets and estimate the tempos. Then, we form a CBSS by taking advantage of OSS and tempo. Next, we perform peak detection by extracting the periodic sequence of beats among all CBSS peaks. The algorithm outperforms state-of-art results in terms of prediction while maintaining comparable and practical computational complexity. The real-time performance is tractable.

## ACKNOWLEDGMENTS

We appreciate the IEEE Signal Processing Society (SPS) to provide us with the opportunity of participating in IEEE SIGNAL PROCESSING CUP 2017 held by IEEE International Conference on Acoustics, Speech and Signal Processing (ICASSP) 2017. The algorithm proposed in this paper was presented as Sharif University team algorithm and received honorable mention as one of the best teams with excellent beat tracking algorithm and annotation. More details about the challenge are available online at [21].